# Strain-induced increase of dielectric constant in EuO thin film

Alireza Kashir[*], Hyeon-Woo Jeong, Woochan Jung, Yoon Hee Jeong, Gil-Ho Lee[*]

Department of Physics, Pohang University of Science and Technology (POSTECH), Pohang, 37673, Republic of Korea

**Abstract**

Recently, lattice dynamics of the highly strained europium monoxide, as a promising candidate for strong ferroelectric-ferromagnet material, applied in the next-generation storage devices, attracted huge attention in the solid-state electronics. Here, the authors investigate the effect of tensile strain on dielectric properties of pulsed laser deposited EuO thin films from 10 to 200 K. A nearly 3% out-of-plane lattice compression is observed as the film thickness was reduced to ~ 8 nm, which could originate from the lattice mismatch between film and a $LaAlO_3$ substrate. The temperature and frequency dependence of capacitance and loss factor of the films reveals the dominant role of electronic and ionic polarization below 100 K. The interdigitated capacitor fabricated on strained film shows almost 50% increase of capacitance compared to the relaxed one, which corresponds to a considerable increase of dielectric constant induced by in-plane tensile strain. Softening of the in-plane polarized transverse optical (TO) phonon modes of EuO lattice due to tensile strain might have the major contribution to this behavior according to the Lyddane–Sachs–Teller relation.

*Corresponding Authors:

lghman@postech.ac.kr
Tel: +82-54-279-2064

kashir@postech.ac.kr

# Introduction

Of the most interesting approaches to generate emergent phenomena in materials, applying strain resulted in a group of physical systems with a range of structural and electronic features, that play a crucial role in the development of solid-state electronics [1-3]. Today, the outstanding progress in the thin film industry in production and characterization enables researchers to produce highly strained materials and control the relaxation processes by using appropriate substrates. This development, in turn, caused the experimental confirmation of recent theoretical predictions and created emergent physical properties which have been hidden behind the bulk materials [4]. It was shown that a proper epitaxial strain adjusts electronic band structure [5], increases transition temperature in superconductors [6], ferromagnetic [7] and also ferroelectric [8-11] materials.

Among the various physical features, dielectric properties and lattice dynamics of materials, which are associated with the structural characteristics, are the most challenging subjects to be studied as a function of strain. The progress in this field may lead to a breakthrough in the development of 21$^{st}$ solid-state physics and therefore, an industrial revolution in energy-storage devices, renewable energy and memory devices.

Choi et al. showed that a biaxial compressive strain enhances the ferroelectric properties of $BaTiO_3$ thin films [9]. Their approach resulted in an increase of ~350 °C in Curie temperature $T_C$ and the remnant polarization $P_r$ was enhanced 250% higher than that of the bulk one. Haeni et al. proved the emergence of ferroelectricity in the quantum para-electric $SrTiO_3$ by applying an appropriate strain [11]. According to their study, the ferroelectric transition temperature arises to the room temperature in a nearly 1% epitaxially strained $SrTiO_3$ thin film grown on $DyScO_3$ substrates. Fennie and Rabe with using first principle DFT calculation predicted that with applying a proper strain coupling between magnetic spins and ferroelectric ordering results in strong multiferroic systems [12]. It was confirmed experimentally by growing $EuTiO_3$ on different substrates to achieve a range of strained systems [13]. Recently, the simplicity of the crystal and electronic structure of rocksalt binary oxides dragged theorists' attention to investigate the dynamical properties of this group of materials under complex condition. Strain-induced ferroelectricity [14-17], pressure-induced Insulator-Metal transition [18-19], Magnetic alignment mechanisms [20], Spin-Phonon coupling at magnetic transition temperature [21] were predicted by first-principle calculations on the binary oxides.

In a pioneering work, Bousquet et al. using first-principles density functional calculations [15] and B. G. Kim by applying soft-mode group theory analysis [14] studied the effect of epitaxial strain on the lattice dynamics of rocksalt binary compounds. It was predicted that the epitaxial strain lowers cubic symmetry to tetragonal and a critical strain causes the condensation of a particular phonon mode resulting in the breaking of inversion symmetry. A substantial softening of the zone-center transverse optical (TO) phonon mode, as the system is under tensile strain, was predicted in their calculation which caused a huge increase of dielectric permittivity. Their approach was applied to the magnetic binary oxides, GdN [16] and EuO [15] and surprisingly it was confirmed that under proper epitaxial strain both oxides show a ferroelectric transition in their preserved magnetic ground state. In recent work, we showed that in a PLD grown strained NiO film, dielectric permittivity increases almost two times higher than that of bulk NiO [22]. Investigation of the effect of strain on dielectric properties of EuO is much more thought-provoking as the larger size of the Eu cation compared to Ni already provides a softer zone-center TO phonon mode of about 199.3 cm$^{-1}$ [23] which is related to the vibrations of Eu and O ions. This vibration is much lower than that of Ni-O bond (384 cm$^{-1}$) [24] resulting in a dielectric constant $\varepsilon_r$ almost 2.5 times higher than NiO. Moreover, the critical strain at which the EuO lattice becomes unstable is considerably lower than that for NiO bulk [15].

EuO belongs to the group of rocksalt binary oxides (space group $F_{m\bar{3}m}$) with a lattice constant of 5.144 Å and a band gap of ~ 1.12 eV [25]. It is the only ferromagnetic binary oxide with rocksalt structure ($T_C$=69 K) [26], having a large local moment on each Eu$^{2+}$ ion from a half-filled 4$f$ band producing a saturation magnetization of 7$\mu_b$ [26]. The effect of strain on the electronic properties of EuO has been studied in theory and experiment [27-28]. Ingle and Elfimov [27] showed that using epitaxial strain the Curie temperature of EuO increases significantly which was experimentally confirmed [25]. Axe obtained the dielectric constant of around 24 for EuO crystal using Infrared reflectivity [23]. By this time, there is no experimentally study on the effect of strain on dielectric properties of EuO thin film. Here, we grow high-quality strained EuO thin films using the pulsed laser deposition technique (PLD) on LaAlO$_3$ substrates and by fabrication of interdigitated electrodes (IDE) the dielectric properties of EuO are studied as a function of strain in the temperature range 10 to 200 K.

# Experiment

## Substrate Selection and Preparation

LaAlO$_3$ (001) (LAO) single crystal (Crystech) was used as a substrate to grow EuO thin films. The relatively low dielectric constant $\varepsilon_r$ of LAO (~ 23) [29] reduces the electric field penetration into the substrate which results in a more precise in-plane dielectric measurement of thin films by fabrication of IDE. In addition, the lattice mismatch between EuO and LAO (~ 4%) enables us to grow strained EuO film provided that the film thickness does not exceed a critical value. LaAlO$_3$ has a rhombohedrally distorted perovskite structure (space group R$\bar{3}$c), which can be described as a pseudocubic with lattice parameters of 3.791 Å [29]. Along the [001] direction, LAO can be viewed as composed of alternating planes of LaO and AlO$_2$ layers. A geometrical consideration would suggest a 45° rotation of EuO relative to LAO during the epitaxial growth. By doing that, a reduction of the misfit from 26% to 4% may achieve [30]. Electrostatic matching also favors this configuration (Fig. 1).

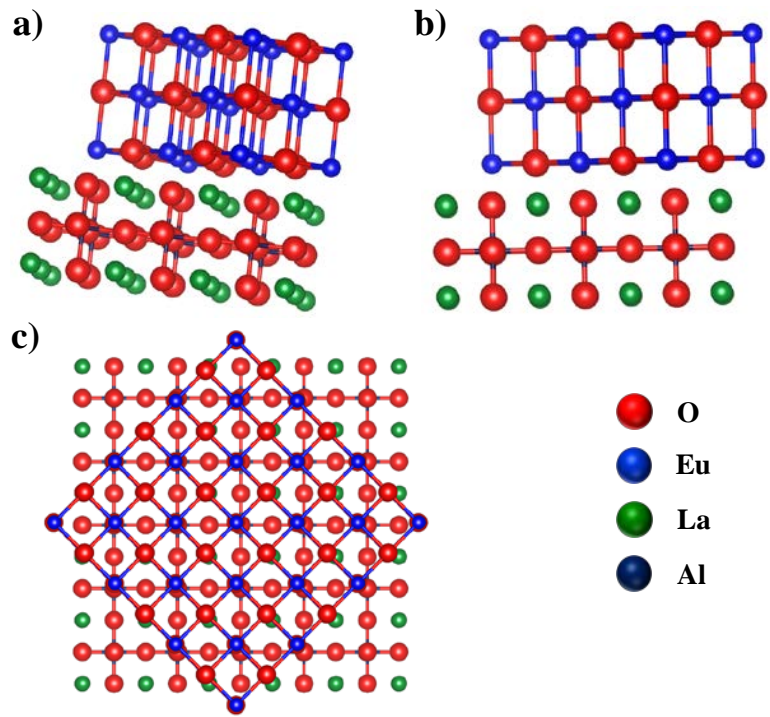

**Fig. 1.** Epitaxial alignment of (001) EuO on pseudocubic (100) LaAlO$_3$. (a) Cross-section view, (b) front view, (c) top view.

To obtain atomically smooth surfaces, LAO substrates were cleaned ultrasonically in acetone for 10 min then in methanol for 10 min to remove contaminants from the top surface. After an ultrasonic soak in deionized water for 5 minutes followed by a 30-second etching process in dilute HCl (pH~4.5), annealing treatment at 1000 °C in the air for 2.5 hours was done [31].

**Film Deposition**

All films were grown by using a polycrystalline Europium metal pellet (99.99% purity) as a target. To ablate the target surface, the laser beam produced by a KrF pulsed laser system with the wavelength of 248 nm (operating at 10 Hz) was focused through an optical lens on the target, generating an energy fluence of ~ 2.5 J/cm$^2$. To obtain a pure EuO phase [32], the substrate was placed at 50 mm above the target. The films were deposited at two different temperatures $T_G$=300 and 350 °C, to achieve a high purity EuO composition without any extra phase. The chamber was pumped to reach the vacuum of 10$^{-8}$ mbar and all films were deposited under the vacuum condition. After film deposition, all grown samples were capped by a 2-nm MgO layer grown under the same condition. Then films were cooled to RT at 10 °C/min.

**Structural Characterization**

The phase analysis and structural evaluation of the deposited materials were investigated by X-ray diffraction technique (XRD) operating at 40 kV and 200 mA using a Cu source as the X-ray generator ($\lambda$=1.541 Å). To study the structural features in detail, a ω rocking around each detected Bragg peak in XRD patterns was done. An X-ray reflectometer (XRR) was used to measure film thickness $t_F$ and the quality of the surface and of the film/substrate interface. To estimate $t_F$, we considered the first five oscillations in the XRR pattern and used equation (1) to obtain four different values.

$$t_F \sim \frac{\lambda}{2} \frac{1}{\theta_{m+1} - \theta_m}, \qquad (1)$$

Where $\lambda$ is the wavelength of X-ray, $\theta_{m+1}$ and $\theta_m$ are the position of *(m+1)-th* and *m-th* interference maximums, respectively. Then $t_F$ was obtained by taking the average of these values and then rounding it to nanometer scale. The surface morphology of the substrates and the deposited samples was investigated using Atomic Force Microscopy (AFM) in dynamic non-contact mode on a Park system XE100 scanning probe microscope.

**Electronic Measurements**

Since the fabrication of a simple capacitor is not an efficient way to measure in-plane electronic properties of a low $\varepsilon_r$ thin film (EuO), we made IDEs [33] with 40 fingers and a small gap (10 µm), which enabled us to investigate the dielectric properties of EuO thin film on a LAO substrate. To define the IDE pattern, we used a commercial conductive polymer layer ("aquaSAVE") on the PMMA spin-coated samples avoiding an accumulation of electrons during e-beam lithography on the insulating film surface. After the patterning process, we gently removed the conductive layer using deionized water. The PMMA was developed using a MIBK solution (MIBK: IPA = 1: 3). Then Ti (5 nm) / Au (35 nm) electrodes were deposited on the pre-made electrode pattern by an e-beam evaporator. Finally, we soaked the sample in acetone for 15 minutes to lift-off the PMMA. (Fig. 2).

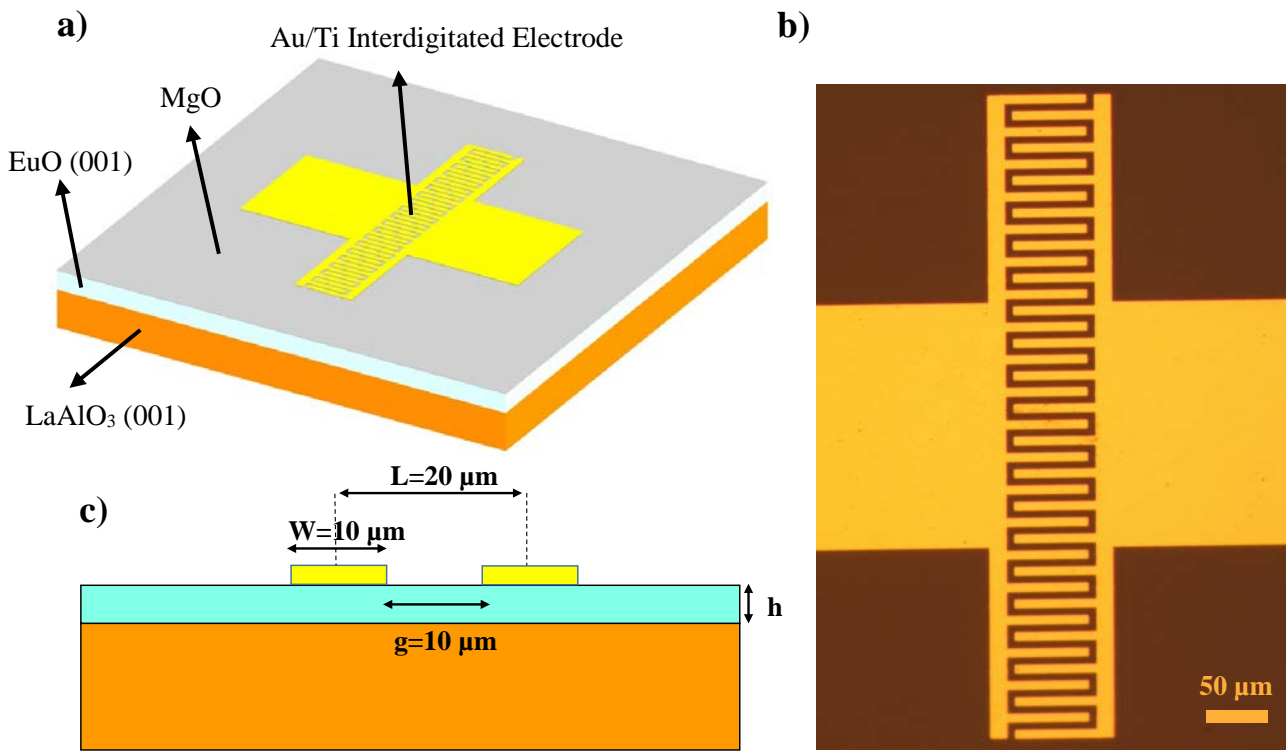

**Fig. 2.** (a) A schematic and, (b) a real image of the Au/Ti interdigitated electrode on EuO films, (c) two-dimensional region surrounding the periodic fingers.

The dielectric properties of IDE capacitors were measured using a Precision LCR meter (Agilent E4980A) at two different frequencies, 1 kHz and 1 MHz, from 10 to 200 K. The temperature was increased using a physical properties measurement system by the rate of 2 K/min during the measurement.

## Result and Discussion

### Structural Characterization

XRD pattern of the film grown at 300 °C showed that it contained two different phases (EuO and $Eu_2O_3$) (Fig. 3). To obtain a single-phase EuO film the $T_G$ was elevated to 350 ºC. Growing a pure EuO film is possible at higher temperatures but to reduce the thermally-driven strain relaxation processes during the growth, the minimum temperature at which a single-phase EuO is achievable was selected as the optimum $T_G$. In this work $T_G$=350 °C appeared to be the proper temperature to grow strained EuO thin film on LAO (001) substrates. A single peak of EuO (002) reveals that the film crystallographic plane grow to match the orientation of the substrate.

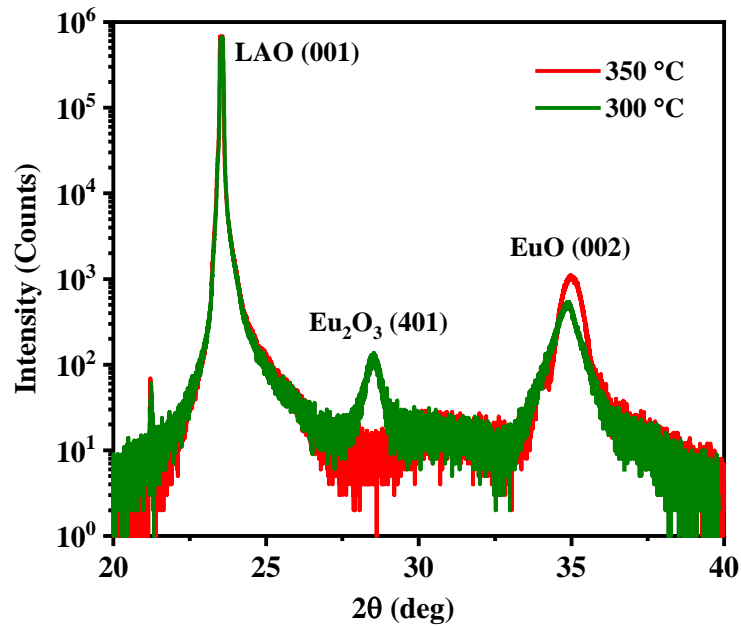

**Fig. 3.** XRD patterns of the deposited thin films on $LaAlO_3$ substrates at different temperatures.

Figure 4.a shows the effect of $t_F$ on the XRD pattern of EuO thin films. Decreasing $t_F$ to 8 nm was accompanied by a right-shift of EuO 002 Bragg peak which indicates an out-of-plane lattice compression. The out-of-plane lattice parameter calculated from the position of 002 Bragg peak

showed ∼ 0.14 Å decrease from bulk value for the thinner film ($a_\perp = 4.992$ Å) which corresponds to an in-plane lattice parameter of 5.222 Å, thus an in-plane tensile strain of 1.5%. The 20-nm film showed an out-of-plane lattice parameter ($a_\perp = 5.133$ Å) almost the same as bulk.

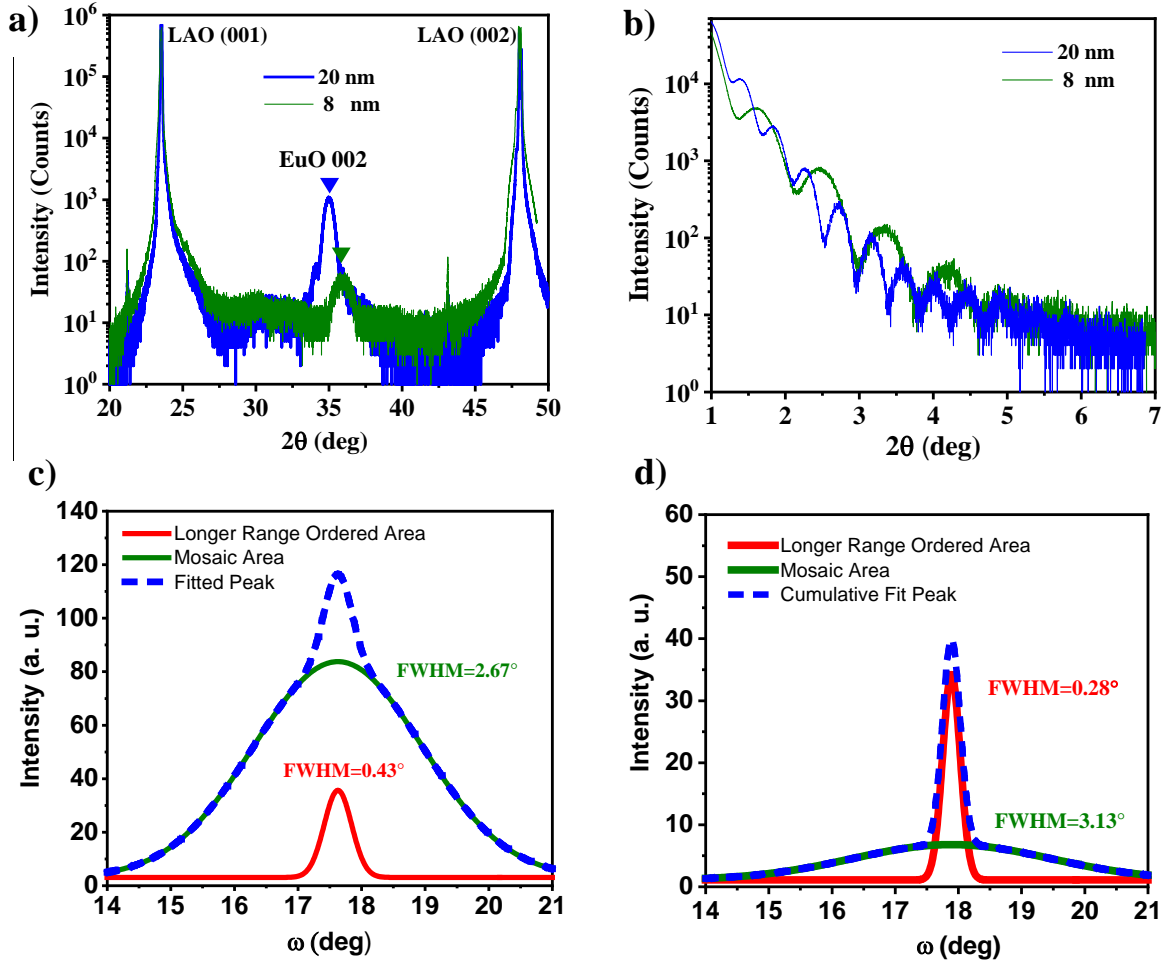

**Fig. 4.** (a) XRD patterns of the EuO thin films with different thicknesses, (b) XRR scans and, ω rocking curve around (002) Brag peak of the EuO film with different thickness c) 20 nm d) 8 nm.

The theoretical lattice mismatch is about 4%, therefore, the relaxation processes started at the initial stages of growth and a partially relaxed film was obtained after the deposition of almost 16 unit cells (∼ 8 nm). The large theoretical lattice mismatch between EuO and LAO caused that the relaxation process starts soon after deposition as the critical thickness for the total relaxation ($h_c$) and theoretical mismatch ($f_{th}$) are inversely proportional (Equation 2) [34].

$$h_c \propto \frac{1}{f_{th}} = \frac{1}{\frac{a_s - a_0}{a_0}} \qquad (2)$$

where $a_s$ and $a_0$ are lattice parameters of the substrate and film, respectively.

The difference between thermal expansion coefficient (TEC) $\alpha_{th}$ of EuO and LAO also causes strain accommodation in the film [35], but its contribution is negligible as the $T_G$ is relatively low and also $\alpha_{th}$ is very close for these two materials ($\alpha_{th, EuO}$=13.2×10$^{-6}$ K$^{-1}$, $\alpha_{th, LAO}$=10×10$^{-6}$ K$^{-1}$). Our calculation showed that the contribution of $\alpha_{th}$ during cooling process is negligible (~ 0.1%).

The X-ray reflectivity patterns of both EuO films showed clear fringes (Fig. 4b), which revealed the high-quality interface between films and substrates and also smooth films' surface. It was confirmed by AFM (Fig. 5). AFM scan on a 3µm × 3µm area of EuO deposited films revealed that the surface roughness did not exceed 1 nm which is almost in the order of the EuO lattice parameter (a = 5.14Å). The film with $t_F$=8 nm showed a clear step and terrace structure. Therefore, during initial stages of growth, the deposited materials seemed to follow the atomic structure of the substrate, and that gradually the clarity of the steps faded. The ω rocking scan around the 002-EuO peak uncovered more information on the crystalline features of grown films. Figure 4.c and 4.d showed that the rocking curve consisted of two different components. A broad component which was due to the mosaic-structured area throughout the film and a narrow component which showed a longer range atomically ordered area [36-38]. The mosaic-structured area is a typical feature of the vacuum grown films as there is no decelerating force to reduce the kinetic energy of the ablated particles resulting in defective structured films. Moreover, the large lattice mismatch between film and substrate (~ 4%) influences the crystalline quality of the deposited material. As $t_F$ decreased, the width of broad peak increased and the sharp peak intensified. The ratio of the broad to sharp components' intensities was negligible for 8-nm film. These changes indicate the spread of the longer-range ordered phase and reduction in mosaic sizes. Setting the growth parameters properly, we obtained a high-quality film with a small fraction of mosaicity to decrease the presence of any possible charges created by defects in EuO structure which consequently affects the dielectric properties of the grown samples.

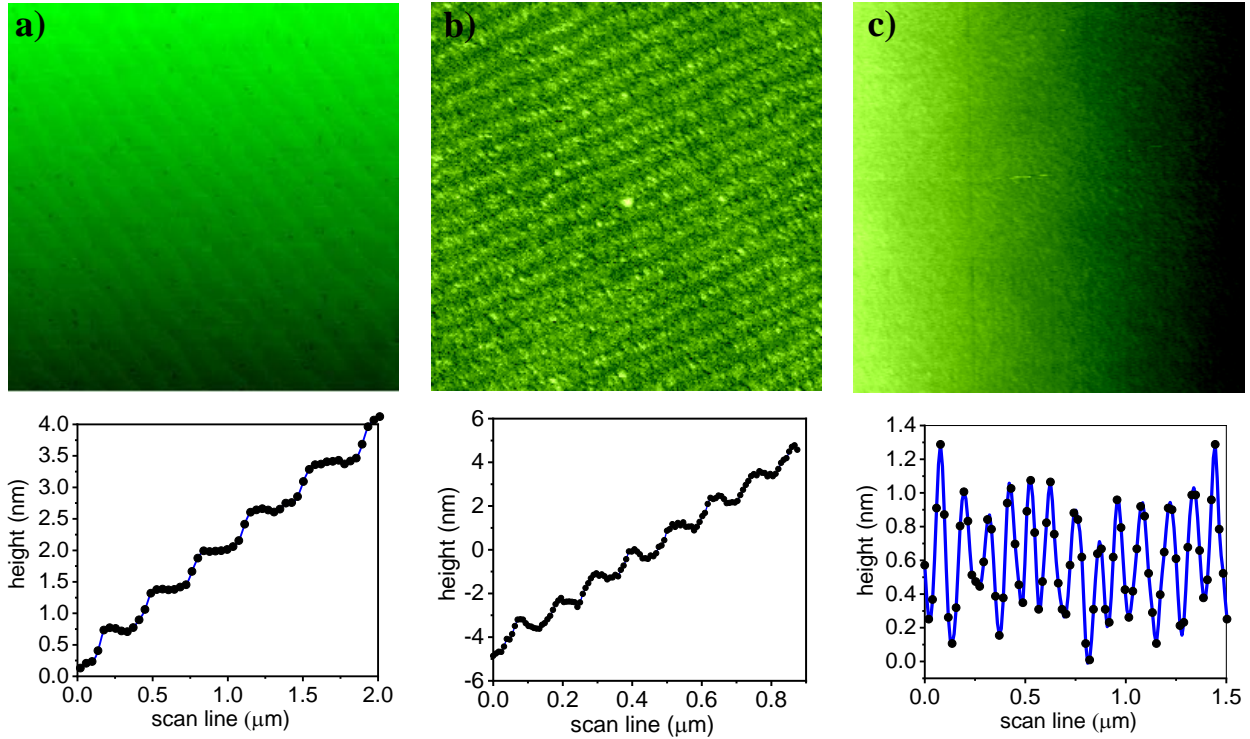

**Fig. 5.** AFM topographic 3µm×3µm images of the (a) LAO (001) substrate after the annealing process, (b) 8-nm, (c) 20-nm EuO films.

**Electronic Measurements**

Figure 6 shows the capacitance of IDE capacitors, fabricated on the 20-nm EuO thin film, as a function of temperature. We measured the capacitance by applying a 2v-AC voltage across the 10-µm gap at two different frequencies, 1 kHz and 1 MHz. The measured capacitances showed almost constant values from 10 to 100 K while above this range a sharp increase was observed which probably indicates the activation of space-charge polarization mechanism as both electronic and ionic mechanisms of polarization, which are activated in rocksalt binary oxides, are not sensitive to the temperature [39]. Moreover, the ω rocking scan around EuO 002 peak indicated the existence of the mosaic area, as mentioned above, which could create space-charges throughout the film structure. These space-charges are thermally-activated species and the activation starts above 100 K. The defective areas, i.e. mosaic boundaries and oxygen vacancies, which can be created during the growth, are the most possible places where space-charges are generated. (As we used high-energy technique for growth).

Increasing the frequency of applied voltage to 1 MHz decreased the high-temperature capacitance while the low-temperature capacitance (below 100 K) remained constant. The strong dependence of capacitance to the applied frequency and temperature above 100 K implied the significant role of space-charge polarization in the dielectric properties of EuO IDE capacitors. On the other side, below 100 K the measured capacitance was independent of the applied frequency and temperature which indicated the dominant contribution of electronic and ionic polarization to the dielectric permittivity of EuO film.

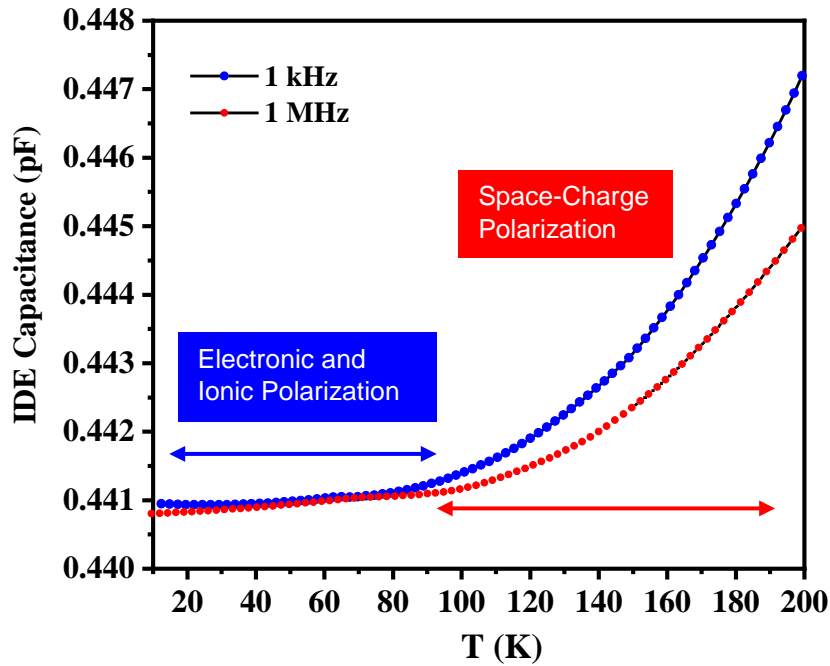

**Fig. 6.** The capacitance of the 20-nm EuO interdigitated capacitor measured at different frequencies.

The loss tangent of both samples showed almost the same behavior under various range of temperature from 10 to 200 K. (Fig. 7). The change in the slope of loss factor at the temperature above 100 K verified the activation of a new polarization mechanism in the samples.

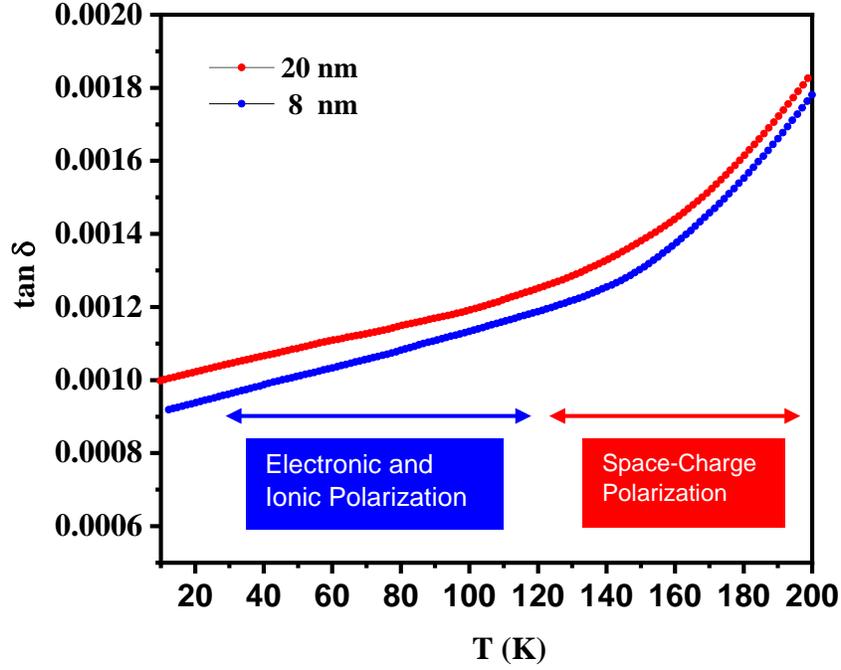

**Fig. 7.** The loss tangent of EuO interdigital capacitors with different thicknesses as a function of temperature.

Figure 8 shows the effect of strain on the measured capacitance of IDE capacitors. The film under 1.5% in-plane tensile strain showed the capacitance almost 50% higher than that of the relaxed one, which implied the effect of in-plane tensile strain on the dielectric constant of EuO.

Farnell et al. [40] used a finite-difference program to evaluate the capacitance of the periodic IDE structure (Equation 3),

$$\epsilon_f = \frac{\left(\frac{C}{K_p} - 1 - \epsilon_s\right)}{1 - exp\left(-\frac{4.6h}{l}\right)} + \epsilon_s \qquad (3)$$

Where $C$ is the measured capacitance per finger length, $h$ is the film thickness, $\epsilon_s$ and $\epsilon_f$ are the dielectric constants of substrate and film, respectively, $l$ is the electrode period and $K_p$ is fitted by a fifth-order polynomial function (Equation 4) modified by Kidner et al. [33],

$$K_p = 137.79 \left(\frac{w}{l}\right)^5 - 320.72 \left(\frac{w}{l}\right)^4 + 288.22 \left(\frac{w}{l}\right)^3 - 120.47 \left(\frac{w}{l}\right)^2 + 28.55 \left(\frac{w}{l}\right) \qquad (4)$$

Where $w$ stands for the electrode width.

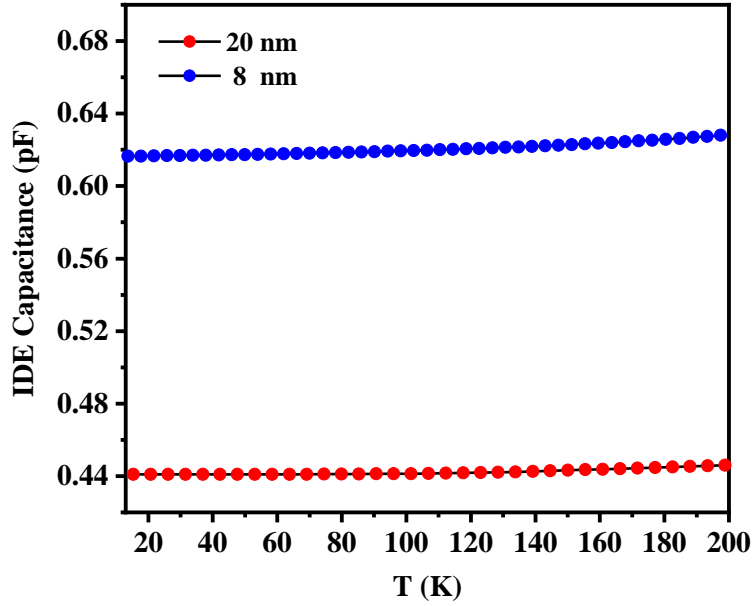

**Fig. 8.** The capacitance of EuO interdigitated capacitor with different thicknesses measured at 1 kHz.

Applying these equations to the IDE pattern showed that 50% rise of the capacitance corresponded to a nearly 5 times increase of the dielectric constant of EuO film which could be due to the strong softening of in-plane TO phonon mode by cause of the 1.5% lattice elongation in x and y-direction. We should point out that to get the perfect fit with their models [33, 40], IDE must have long fingers compared to the finger width (Finger width/Finger length ~ 0). In our work the ratio is ~ 0.1, then some errors could happen in calculation of absolute values. Therefore, we did not report absolute values of $\varepsilon_r$ here.

Recently, Pradip et al. deposited ~ 2% strained EuO film on YAlO$_3$ (110) substrate. Employing nuclear inelastic scattering on the Mössbauer-active isotope $^{151}$Eu and first-principles theory they carried out a systematic lattice dynamics study of ultrathin epitaxial EuO (001). They observed the number of phonon states at low energy increases significantly by an increase of the tensile strain [41].

Bousquet et al. [15] using first-principles density functional calculations, predicted the increase of $\varepsilon_r$ in EuO under epitaxial tensile strain. According to their calculation under in-plane biaxial tensile strain the triple-degenerate cubic TO phonon mode at zone-center (Γ point) with F$_{1u}$ symmetry splits to a single non-degenerate A$_{2u}$ mode polarized along the [001] axis, and a two-fold degenerate E$_u$ mode polarized along [100] or [010] axis, resulting in a phase transition to tetragonal structure (space group $I_{4/mmm}$) (Fig. 9). The increase of in-plane tensile strain decreases the frequency of the two-fold degenerate E$_u$ phonon mode, resulting in an increase in the in-plane $\varepsilon_r$ of EuO thin film. The tensile strain (increasing bond length) decreases the electron overlap between anion and cation, resulting in a weak bonding between them and subsequently a softening of the TO phonon mode which is a basic component in the calculation of dielectric permittivity of a rocksalt binary oxide according to the Lyddane–Sachs–Teller relation (Equation 5) [42].

$$\frac{\epsilon(0)}{\epsilon(\infty)} = \prod_j \frac{\omega_{Lj}^2}{\omega_{Tj}^2} \qquad (5)$$

Where $\omega_L$ and $\omega_T$ are the frequencies of the longitudinal and transverse optical phonon modes, respectively, $\varepsilon(0)$ is the low-frequency dielectric permittivity and $\varepsilon(\infty)$ is the high-frequency limit for electronic dielectric permittivity. Decreasing $\omega_T$ due to the tensile strain is accompanied by an increase in the low-frequency dielectric constant.

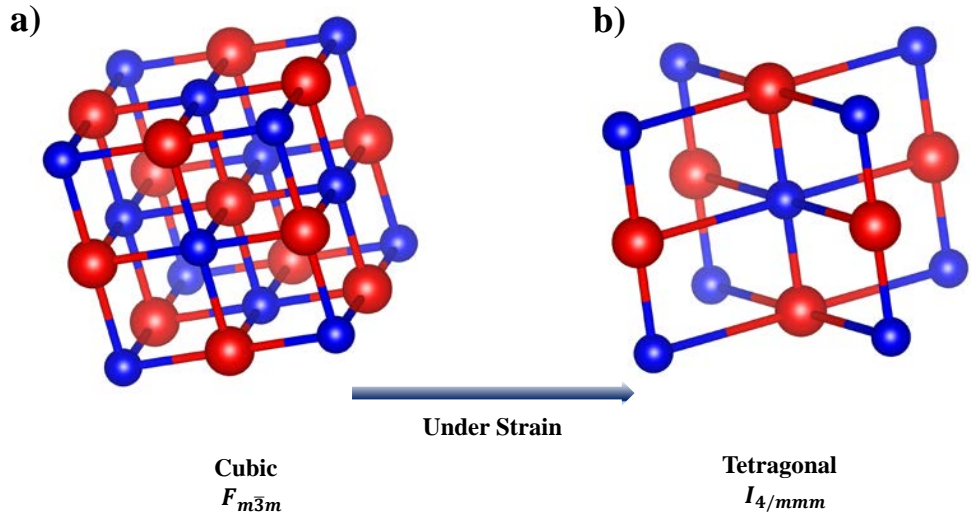

**Fig. 9.** The structural change of EuO under biaxial strain, (a) a cubic rocksalt structure with a triple degenerate TO phonon mode along x, y and z-axis. (b) A tetragonal structure with a single non-degenerate A$_{2u}$ mode polarized along the z-axis and a twofold degenerate E$_u$ mode polarized along x and y-axis.

# Conclusion

Dielectric properties of the PLD grown EuO thin films have been studied as a function of strain and temperature. The XRD and AFM results revealed that atomically long-range ordered EuO films are achievable using the PLD technique. A nearly 3% out-of-plane lattice compression in the 8-nm EuO film on LAO substrate (corresponds to ~ 1.5% in-plane tensile strain) was observed which is due to the lattice mismatch of film and substrate. The in-plane capacitance measurements of the strained film showed a dielectric constant almost 5 times higher than that of the relaxed one. This behavior was predicted based on the investigation of the lattice dynamics of rocksalt binary compounds. This result implied the substantial effect of tensile strain on the softening of TO phonon mode of EuO, which is a basic component in the dielectric properties of rocksalt binary oxides according to the Lyddane–Sachs–Teller relation.

# Acknowledgments

This work was supported by the National Research Foundation (NRF) of Korea (2015R1D1A1A02062239 and 2016R1A5A1008184) funded by the Korean Government.